\documentclass{svjour3}                     
\smartqed  
\usepackage{graphicx}
\usepackage{latexsym,amsbsy,amssymb,amsmath}
\usepackage[T1]{fontenc}
\journalname{Few-Body Systems (FB20)}
\begin{document}

\title{Stability of multiquarks in an improved flip-flop model of confinement\thanks{Presented at the 20th International IUPAP Conference on Few-Body Problems in Physics, 20 - 25 August, 2012, Fukuoka, Japan}
}


\author{J. Vijande \and A. Valcarce \and J.-M. Richard}

\institute{J. Vijande \at
              Dpto. de F\' \i sica At\'omica, Molecular y Nuclear - IFIC,
Universidad de Valencia - CSIC, Spain\\
              \email{javier.vijande@uv.es}           
           \and
           A. Valcarce \at
              Dpto. de F{\'\i}sica Fundamental,
Universidad de Salamanca, Spain
           \and
           J.-M. Richard \at 
Laboratoire de Physique Subatomique et Cosmologie,
 Universit\'e Joseph Fourier--INPG--IN2P3-CNRS, France}

\date{Received: date / Accepted: date}

\maketitle

\begin{abstract}
We review some recent studies on the string model of confinement inspired by the strong-coupling regime of QCD and its application to exotic multiquark configurations. This includes two quarks and two antiquarks, four quarks and one antiquark, six quarks, and three quarks and three antiquarks with a careful treatment of the corresponding few-body problem.

\end{abstract}


The physics of multiquark systems remains rather delicate. On the experimental side, 
most of the candidates which have been announced are not yet confirmed  \cite{Nak2010}. Moreover, the search has been focused mainly on states with hidden flavor, which might be either genuine multiquarks or ordinary hadrons strongly coupled to their decay channels.
Other sectors such as double charm mesons have not been much explored. 

As for the theory, the main question is how to extrapolate quark models adapted to mesons and baryons toward the multiquark sector. Then it remains to solve carefully the few-quark problem. Many configurations lie at the edge between binding and instability, so the wave function should incorporate both short-range components where the quark interaction takes place and long-range components which include the clusterized part of the wave function.

\begin{figure}
\includegraphics[width=.2\textwidth]{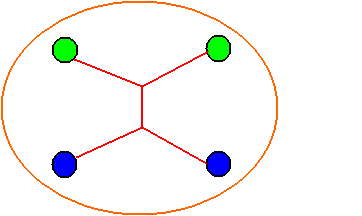}\
\includegraphics[width=.2\textwidth]{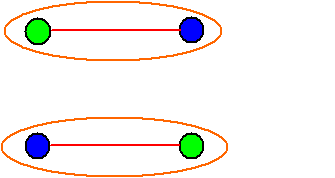}\
\includegraphics[width=.2\textwidth]{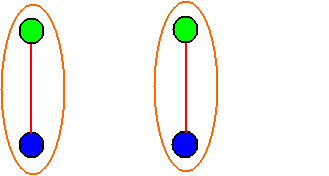}\qquad
\includegraphics[width=.2\textwidth]{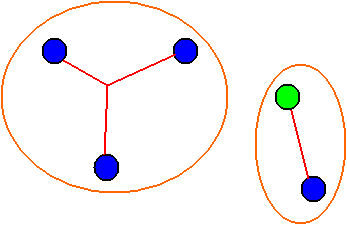}
\\[.4cm]
\includegraphics[width=.2\textwidth]{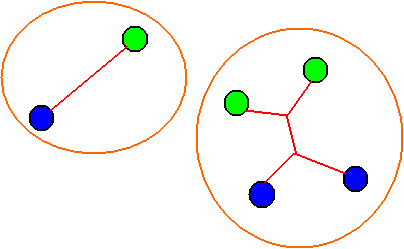}\quad
\includegraphics[width=.15\textwidth]{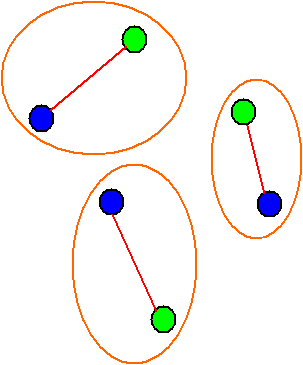}\qquad
\includegraphics[width=.2\textwidth]{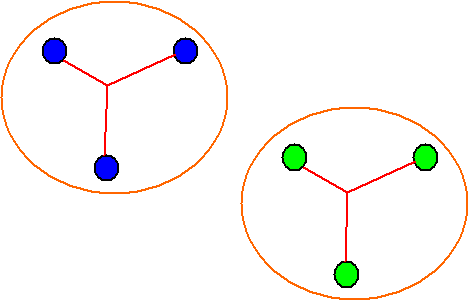}\qquad
\includegraphics[width=.15\textwidth]{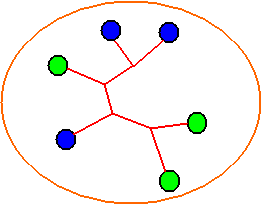}
\caption{\label{fig1}A few examples of flip-flop terms and connected Steiner trees. From left to right,top to bottom: the three contributions to the tetraquark, the flip-flop term of the pentaquark (to be minimized by permuting the quarks), and some disconnected and connected contributions to the $(Q^3\bar q^3)$ confinement.}
\end{figure}

In the studies summarized here, we focused on the role of confinement, with a string model inspired by \cite{Len1985,Car1991}. The linear potential corresponds to the minimal length for the gluon flux tube linking the quark to the antiquarks. For baryons, it becomes the $Y$-shape interaction which minimizes the cumulated length of the flux tubes. For tetraquarks and beyond, the potential is mainly of the flip-flop terms with an interaction inside each cluster and a minimization over all possible cluster decompositions. There are  connected Steiner diagrams which link all quarks and antiquarks, see Fig.~\ref{fig1}. 

In Refs.~\cite{Vij2007,Ric2009,Vij2012}, we solved the tetraquark, pentaquark and hexaquark problems with this confining potential alone, for systems whose quarks are assumed to be different, even in the case where they bear the same mass. This means that are neglected altogether antisymmetrization, relativistic effects, short-range corrections, spin-dependent effects, etc. We are now studying the interplay with chromomagnetic forces and the role of antisymmetrization.

This string model replaces the ``color-additive'' ansatz that was widely used in the early days of hadron spectroscopy and reads

\begin{equation}\label{intro:eq:l-l-rule}
V(\vec{r}_1,\ldots)=-{3\over 16}\sum_{i<j} \tilde{\lambda}_i.\tilde{\lambda}_j v(r_{ij})~.
\end{equation}
%
For baryons, the color-additive model gives the well-known ``1/2'' rule between the quark--quark and the quark--antiquark potentials  \cite{Stan1980}, while the string model leads to~\cite{Tak2000} 
\begin{equation}\label{intro:eq:Y-pot}
Y(\vec{r}_1,\vec{r}_2,\vec{r}_3)= \lambda \min_k (r_{k1}+r_{k2}+r_{k3})~,
\end{equation}
reminiscent of the celebrated problem of Fermat and Torricelli, later generalized as the Steiner problem \cite{Steiner}.

For tetraquarks, the potential is the minimum of the flip-flop and connected Steiner-tree terms. The former is the minimum of the two possible configurations made of two mesons, and the later is a generalization of the $Y$-shape interaction, with optimization with respect to two junction points. More precisely,
\begin{equation}\label{inter:eq:str}
\begin{aligned}
V_{s}&=\min(V_{f},V_{b})~,\\
V_{f}&=\lambda \min( r_{13}+r_{24},r_{23}+r_{14}),\quad
V_{b}=\lambda \min_{k,\ell} (r_{1k}+r_{2k}+r_{k\ell}+r_{\ell 3}+r_{\ell 4})~.
\end{aligned}
\end{equation}
%
%
For a detailed discussion see Ref.~\cite{Vij2007}.

\begin{table}[t]
\caption{\label{resu:tab:Mm} Four-quark energies, $E_4$, for the different confinement models described in Eq. (\ref{inter:eq:str}) 
compared to their threshold, $T_4$, as a function of the mass ratio $M/m$.}
\begin{tabular}{c|cccc||cc}
$M/m$	& \multicolumn{3}{c}{$E_4(QQ\overline{q}\overline{q})$}	& $T_4(QQ\overline{q}\overline{q})$ & $E'_4(Q\overline{Q}q\overline{q})$ & $T'_4(Q\overline{Q}q\overline{q})$\\ 
	& $V_f$	& $V_b$	& $V_s$	&  & $V_f$  &\\ 
\hline
1	& 4.644	& 5.886	& 4.639	& 4.676  &  4.644 & 4.676 \\
2	& 4.211	& 5.300	& 4.206	& 4.248  & 4.313  & 4.194\\
3	& 4.037	& 5.031	& 4.032	& 4.086  & 4.193  & 3.959 \\
4	& 3.941	& 4.868	& 3.936	& 3.998  & 4.117  & 3.811\\
5	& 3.880	& 4.754	& 3.873	& 3.942  & 4.060  & 3.705\\
\end{tabular}
\end{table}
The results are shown in Table \ref{resu:tab:Mm}. Clearly, as $M/m$ increases, a deeper binding is obtained for the flavor-exotic 
$(QQ\bar{q}\bar{q})$ system. For the hidden-flavor $(Q\overline{Q}q\bar{q})$, however, the stability deteriorates, and with our variational approximation, for 
$M/m\gtrsim 1.2$, the system becomes unbound with respect to its lowest threshold $(Q\overline{Q})+(q\bar{q})$.


The pentaquark potential of Ref.~\cite{Ric2009} is similarly a combination of flip-flop and connected terms
\begin{equation}\label{eq:VP}
V_P=\min(V_\text{ff},V_\text{St})~,\qquad
V_\text{ff}=\min_i\left[r_{1i}+ V_Y(\vec{r}_j,\vec{r}_k,\vec{r}_\ell)\right]~,
\end{equation}
with $\{i,j,k,\ell \}$ suitably permuted,
and $V_\text{St}$ denoting the connected Steiner trees. 

The calculation was performed using a hyperspherical approximation and a Gaussian expansion. 
Both methods agree that systems made of four light quarks and one light antiquark are clearly below the lowest meson-baryon threshold. This is the opposite to what 
occurs when a pair-wise interaction is considered. 

If the calculation is repeated in the case of four unit masses and an infinitely massive antiquark or quark,  the 
pairwise model gives an energy well above the threshold for both $(\overline{Q}qqqq)$ and $(\bar{q}qqqQ)$. 
On the other hand, the flip--flop potential gives 
$E_\text{ff}(\overline{Q}qqqq)-E_\text{th}(\overline{Q}qqqq)\simeq -0.114$, and $E_\text{ff}(\bar{q}qqqQ)-E_\text{th}(\bar{q}qqqQ)\simeq -0.2777$, 
which is sufficient to demonstrate their stability.


For the $(q^6)$ configurations, there are again two types of diagrams: flip-flop and connected Steiner trees. The potential is minimized with respect to all permutations.                                  
For the $(q^3\bar q^3)$ states, there are several possibilities: flip-flop with either a baryon and an antibaryon, or three mesons, or a meson and a tetraquark, and 
also some connected diagrams with four or more junctions. The dynamics is dominated by the flip-flop terms, while the connected diagrams with $Y$-shape junctions play a minor role for binding. Moreover, the dynamics of baryon 
is qualitatively similar with a pair-wise potential $\sum r_{ij}/2$ and the $Y$-shape model, this leading to easier computations without spoiling the spirit of the model.
 For a more detailed discussion see Ref.~\cite{Vij2012}.


The results are shown in Table \ref{results}. For the ``dibaryon'', $(Q^3q^3)$, the system is found stable against dissociation into two baryons. However,  
the stability deteriorates when the mass ratio increases. The behavior is reasonably linear and therefore the  limit where the system becomes unbound can be estimated 
to be of the order of $M/m\approx8-10$. Hence a triple-charm dibaryon is predicted but not a triple-beauty one.

In the case of $(Q^3\bar q{}^3)$, this is more delicate. For $M=m=1$, the results in Table~\ref{results} suggest there is a 
shallow bound state below the lowest threshold. This means that the effective interaction between the $(q\bar q)$ mesons is
attractive. Now, as $M/m$ further increases, the $(Q^3)+(\bar q{}^3)$ threshold will become degenerate with the lowest mesonic threshold. This will favor binding, as the six-body wave 
function will contain two different decompositions into clusters with relative motion that will interfere to improve binding.  However, for even larger values of the mass ratio 
$M/m$, no multiquark configuration can acquire enough binding to compete with the compact $(Q^3)$, and the system becomes unstable against rearrangement into $(Q^3)+(\bar q^3)$. 
Perhaps, some metastability could be observed with respect to some higher threshold. 

\begin{table}[t]
\caption{\label{results} $(Q^3q^3)$ and $(Q^3\bar q{}^3)$ variational energies
$E$, compared to their threshold energy $T$, as a function of the mass ratio
$M/m$. $\Delta=E-T$ is the energy
difference.}
\begin{tabular}{c|ccc||ccc}
$M/m$		& $E(Q^3q^3)$	& $T(Q^3q^3)$	& $\Delta(Q^3q^3)$	& $E(Q^3\bar q{}^3)$	& $T(Q^3\bar q{}^3)$	& $\Delta(Q^3\bar q{}^3)$\\
\hline
1		& 7.237		& 7.728		& -0.491	&
6.981		& 6.981		& +0.000	\\
2		& 6.524		& 6.929		& -0.405	&
6.314		& 6.335		& -0.021	\\
3		& 6.209		& 6.543		& -0.334	&
6.030		& 6.079		& -0.049	\\
4		& 6.014		& 6.298		& -0.294	&
5.868		& 5.940		& -0.072	\\
5		& 5.890		& 6.123		& -0.233	&
5.762		& 5.852		& -0.090	\\
\end{tabular}
\end{table}


Let us summarize and suggest some possible further studies.

\textsl{1.} The string model of confinement which combines flip-flop and connected  flux tubes of minimal length gives more attraction than the additive pairwise model with color factors that was used in early multiquark calculations.

\textsl{2.} This potential  is flavor independent. By changing the constituent  masses in the kinetic-energy part of the  hamiltonian, one can modify the binding. For tetraquarks, $(QQ\bar q\bar q)$ is more stable when the mass ratio $M/m$ increases, while $(Q\overline{Q}q\bar{q})$ becomes unstable. For pentaquarks the calculation has been restricted to five equal and finite masses, or one infinitely massive constituent surrounded by four equal masses. The property of stability was found to survive for both $(\overline{Q}qqqq)$ and $(\bar{q}qqqQ)$. In the six-quark case the binding energy decreases for $(Q^3 q^3)$, while for $(Q^3 \bar q{}^3)$, it first increases and then decreases. In the limit of large $M/m$, no six-body configuration can compete with the deep binding of $(Q^3)$
that enters the lowest threshold.

\textsl{3.} These potentials can be seen as a Born--Oppenheimer limit. 
Our aim is to reformulate the interaction as an operator in color space, to go beyond the adiabatic limit and include the constraints of antisymmetrization.

This work has been partially funded by the Spanish Ministerio de Educaci\'on y Ciencia and EU FEDER under Contracts No. FPA2010-21750 
and AIC-B-2011-0661, and by the Spanish Consolider-Ingenio 2010 Program CPAN (CSD2007-00042).
The participation of JMR to this Conference was partially supported by the French-Japanese cooperation program in Nuclear Physics.


\end{document}